\documentclass[conference, letterpaper]{IEEEtran}
\usepackage[top=0.7in, bottom=1.0in, left=0.65in, right=0.65in]{geometry}

\usepackage[short]{optidef} 

\usepackage{xcolor}

\definecolor{ao}{rgb}{0.0, 0.6, 0.0}
\definecolor{ppl}{rgb}{0.4940 0.1840 0.5560}
\usepackage{pgfplots}
\usepackage[caption=false]{subfig}
\pgfplotsset{compat=1.18}
\usepackage{tikz}
\usetikzlibrary{calc}
 
\newcommand{\gettikzxy}[3]{%
  \tikz@scan@one@point\pgfutil@firstofone#1\relax
  \edef#2{\the\pgf@x}%
  \edef#3{\the\pgf@y}%
}

\usepackage{lettrine}

\usepackage{graphics,
	psfrag,
	epsfig,
	amsthm,
	cite,
	amssymb,
	url,
	dsfont,
	balance,
	enumerate,
	color,
	setspace
}

\usepackage{multirow}
\usepackage{algorithm} 
\usepackage{algorithmicx}
\usepackage{algpseudocode}
\usepackage{xcolor}
\usepackage{soul}
\usepackage{amsmath}

\usepackage{epstopdf}

\newcommand{\ignore}[1]{}

\usepackage{comment}
\usepackage{derivative}

\newcommand{\eref}[1]{(\ref{#1})}

\newcommand{\cref}[1]{Constraint~\ref{#1}}

\IEEEoverridecommandlockouts
\usetikzlibrary{spy,backgrounds}
\usetikzlibrary{shapes, arrows, positioning}

\usepackage{siunitx}

\usepackage{mathtools}
\usepackage{color}
\usepackage[utf8]{inputenc}

\begin{document}
\IEEEoverridecommandlockouts

\title{Joint Movable Antenna Positioning and RIS Partitioning for Sum-Rate Maximization  



 
}

\author{\IEEEauthorblockN{ Mohammed Saif}
\IEEEauthorblockA{Electrical, Computer, and Biomedical Engineering, Toronto Metropolitan University, Toronto, ON, Canada  \\
Email: mohammed.saif@torontomu.ca}
}

\maketitle

         





\begin{abstract}

This paper investigates the utility of the movable antenna (MA) and reconfigurable intelligent surface (RIS) framework for  downlink wireless communications. In the considered scenario, a base station (BS) is equipped with two sub-arrays of MAs transmits signals to the users via the RIS. By jointly exploiting the antenna-positioning flexibility of MAs and the RIS element selection capability, the proposed joint MA-RIS framework introduces additional design degrees of freedom to enhance desired signals and mitigate inter-user interference, thereby maximizing the network sum-rate. To this end, we formulate a joint optimization problem involving MA positioning, sub-array beamforming, and RIS element selection, subject to the minimum antenna separation and transmit power constraints. The resulting problem is highly non-convex and challenging to solve directly. To address this issue, an alternating optimization framework is developed that decomposes the problem into three tractable subproblems. Specifically, zero-forcing beamforming is employed for transmit beamformer design, a low-complexity one-dimensional search is derived for RIS element selection, and the MA positioning problem is solved using block coordinate descent (BCD) and convex optimization techniques implemented via CVX. Simulation results demonstrate that the proposed joint MA-RIS framework significantly improves the achievable sum-rate compared with conventional fixed MAs  and benchmark schemes with random configurations.


\end{abstract}
 

\section{Introduction}
The rapid evolution of wireless communication systems has driven significant interest in the development of sixth-generation (6G) networks \cite{10560514}. Future 6G systems are expected to support stringent performance requirements, including higher data rates, enhanced reliability, and improved spectral efficiency, thereby necessitating the exploration of advanced transmission technologies \cite{10854532, 10555049}. In this context, multi-antenna communication techniques have emerged as one of the most promising solutions, owing to their ability to exploit spatial multiplexing and enhance both spectral and energy efficiency.

Conventional multi-antenna architectures rely on antennas deployed at fixed positions, which limits the spatial degrees of freedom available for achieving diversity and multiplexing gains \cite{10555049}. Recently, a new frontier technology known as movable antennas (MAs) has emerged as a promising approach to enhance the communication channel between the base station (BS) and users by enabling six-dimensional adjustments, including three-dimensional positions \cite{10753482, 10286328, 11395099}.
Unlike traditional fixed antenna deployments, MAs can dynamically adapt their positions to better align with channel conditions. As a result, the MA paradigm can achieve improved performance using the same number of antenna elements. Moreover, it is compatible with most existing beamforming and antenna techniques, allowing seamless integration into current wireless systems \cite{10286328}.

Despite the advantages of MAs in enhancing spectral efficiency and communication reliability, blockage of the direct communication link can significantly degrade their performance, leading to coverage holes and increased outage probability. To address this limitation, reconfigurable intelligent surfaces (RISs) have emerged as an effective solution for improving coverage by dynamically shaping the wireless propagation environment \cite{8811733}. An RIS typically consists of a planar array with a large number of passive reflecting elements. By adaptively adjusting the reflection coefficients of these elements, the RIS can intelligently control signal propagation \cite{9756313}. In particular, through optimized reflection design, RISs can steer incident signals toward intended users, thereby enhancing channel conditions and mitigating coverage blind spots \cite{10458024, 8811733}. Owing to their largely passive operation—with power consumption mainly associated with control circuitry—RISs provide a cost-effective, low-maintenance, and easily deployable solution. Motivated by these advantages, integrating MAs with RIS-assisted propagation control offers a promising approach for enabling adaptive spatial control, high spectral efficiency, and reliable coverage in future 6G systems.

Existing works utilized the integration of RIS and MA for sum-rate maximization \cite{10437926}, sensing and communications \cite{11033708}, and coverage \cite{10354003}, however, none of these works considered RIS partitioning and MA sub-array design. In this paper, we design a system framework of  MA sub-arrays that communicate with users via RIS partitioning with a joint optimization of RIS partitioning, MA sub-array beamforming, and their positions. The main contributions of this paper are summarized as follows:
\begin{itemize}

\item We develop a joint optimization MA-RIS framework for sum-rate maximization by optimizing the transmit beamforming of the sub-arrays, the positions of MAs, and RIS partitioning, subject to power budget and mutual coupling constraints.

\item To address the non-convexity of the formulated problem, we propose an alternating optimization scheme. Specifically, the transmit beamforming subproblem is reformulated using the Karush--Kuhn--Tucker (KKT) conditions to obtain a tractable solution. The RIS partitioning subproblem is reduced to a one-dimensional search space, while the MA position optimization subproblem is handled using a block coordinate descent (BCD)-based approach to obtain a convex reformulation.

\item Numerical results demonstrate the effectiveness of the proposed joint MA-RIS framework, showing significant performance gains over benchmark schemes.

\end{itemize}
The remainder of this paper is organized as follows. Section II presents the system model. In Section III, the joint MA-RIS framework is developed. Section IV discusses the simulation results, and Section V concludes the paper

\begin{figure}  
\begin{center}
\includegraphics[width=0.99\linewidth, draft=false]{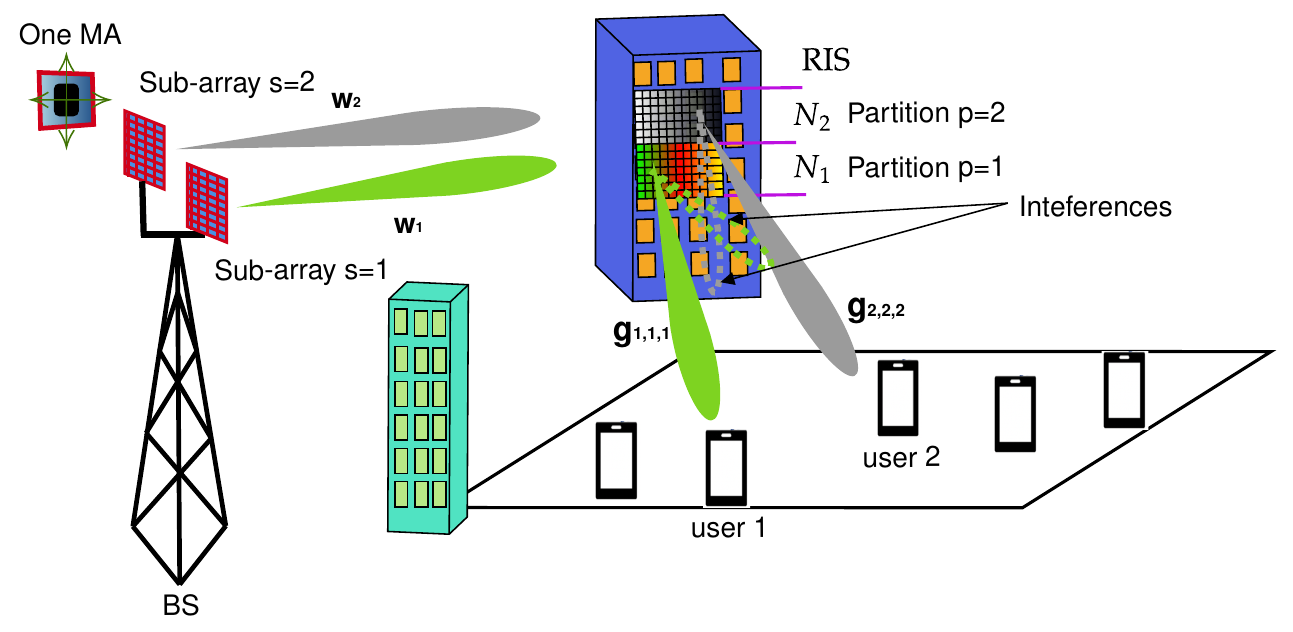}
\caption{System model.}
   \label{fig1}
\end{center}

\end{figure}

\section{System Model}\label{S}
\subsection{Channel Model}
As shown in Fig.~\ref{fig1}, a downlink RIS-assisted network is
considered where a BS equipped with $M$ MAs communicates with a set of $\mathcal D$ users with the help of RIS.  In dense urban environments, where the direct links between the BS and users are obstructed, communication is established exclusively through the RIS. 
Similar to \cite{11033708, asif2026channeluncertaintyawarerobustbeamforming}, we adopt a planar far-field channel model, where all transmit/receive paths of a given channel are assumed to share the same angle-of-departure (AoD), angle-of-arrival (AoA), and path gain. However, the phase of each path response varies across different MA positions due to their distinct spatial locations.

The RIS is equipped with \(N=N_hN_v\) reflecting elements arranged in a uniform planar array (UPA), where \(N_h\) and \(N_v\) denote the numbers of RIS elements along the horizontal and vertical dimensions, respectively. The RIS can reflect the signal of the BS to multiple users through virtual partitioning representation. The set of users that exploit the RIS is denoted as $\mathcal D_\text{RIS}$, which also represents the set of partitions in the RIS and $ D_\text{RIS}=|\mathcal D_\text{RIS}|$. 
For simplicity of analysis, this paper considers that the BS is equipped with $2$ antenna sub-arrays sending their signals to $2$ users via the RIS. The total number of BS MAs is $M = \sum_{s=1}^{2} M_s$, where $M_s$ is the number of MAs in the $s$-th sub-array and $M_1=M_2=M/2$, and $M$ is an even number. The RIS is partitioned into two portions, where each portion is configured to passively beamform the signal of one sub-array to one user (i.e., $\mathcal D_\text{RIS}=\{\text{user 1}, \text{user 2} \}$ as illustrated in Fig. \ref{fig1}. Thus, we consider 
two aligned signals $\text{sub-array} ~1 \xrightarrow{\text{RIS}_{p=1}} \text{user}~{1}$ and $\text{sub-array} ~2 \xrightarrow{\text{RIS}_{p=2}} \text{user}~{2}$, where the triple $(1, 1, 1)$ represents sub-array $s=1$, RIS partition $p=1$, and user $d=1$. For $\text{sub-array} ~1 \xrightarrow{\text{RIS}_{p=1}} \text{user}~{1}$, the number of RIS elements allocated to reflect the intended signal is denoted by $N_{1}=\lceil\rho_{1}N\rceil$, where $\rho_{1} \in [0, 1]$ is the RIS allocation factor such that $\sum_{p=1}^2 \rho_{p}  \leq 1$ and $\sum_{p=1}^2 N_{p}=N$. The RIS partitioning vector is $\boldsymbol{\rho} = [\rho_p, \rho_{p'}]$.

 We define the positions of $M_s$ MAs of the $s$-th sub-array as
$\mathbf{p}_s = [\mathbf{p}_{s,1}, \mathbf{p}_{s,2}, \ldots, \mathbf{p}_{s,M_s}] \in \mathbb{R}^{2 \times M_s}$,
where $\mathbf{p}_{s,m} = [x_{s,m}, y_{s,m}]^T \in \mathcal{C}_s$ denotes the 2D position of the $m$-th MA, $\forall m \in \mathcal{M}_s = \{1,2,\ldots,M_s\}$, and $\mathcal{C}_s$ represents the transmit region of the $M_s$ MAs at the BS. Let $M_{s,t}$ and $M_{s,r}$ denote the numbers of transmit and receive paths, respectively. The elevation and azimuth angles at the BS are denoted by $\varphi_e^{i} \in [0,\pi]$ and $\varphi_a^{i} \in [0,\pi]$, for $1 \leq i \leq M_{s,t}$, while those at the RIS are denoted by $\phi_e^{j} \in [0,\pi]$ and $\phi_a^{j} \in [0,\pi]$, for $1 \leq j \leq M_{s,r}$. The difference in signal propagation distance for the $i$-th transmit path between $\mathbf{p}_{s,m}$ and the reference origin is given by \cite{ asif2026channeluncertaintyawarerobustbeamforming}
\begin{equation}
\chi\big(\mathbf{p}_{s,m}, \varphi_e^{i}, \varphi_a^{i}\big)
= x_{s,m} \sin(\varphi_e^{i}) \cos(\varphi_a^{i}) + y_{s,m} \cos(\varphi_e^{i}).
\end{equation}

For the BS--RIS link, the field response vector (FRV) of the MA at $\mathbf{p}_{s,m}$ is given by $\mathbf{f}(\mathbf{p}_{s,m}) =
\left[
e^{j \frac{2\pi}{\lambda} \chi(\mathbf{p}_{s,m}, \varphi_e^{1}, \varphi_a^{1})},
\ldots,
e^{j \frac{2\pi}{\lambda} \chi(\mathbf{p}_{s,m}, \varphi_e^{M_{s,t}}, \varphi_a^{M_{s,t}})}
\right]^T \in \mathbb{C}^{M_{s,t} \times 1}$,
where $\frac{2\pi}{\lambda}\chi(\mathbf{p}_{s,m}, \varphi_e^{i}, \varphi_a^{i})$ denotes the phase difference of the $i$-th transmit path between the MA position $\mathbf{p}_{s,m}$ and the reference origin, and $\lambda$ is the carrier wavelength. Accordingly, the field response matrix (FRM) for the BS--RIS communication link across all $M_s$ MAs of the $s$-th sub-array is given by
\begin{equation}
\mathbf{F}_\text{BS}(\mathbf p_s) = 
\left[
\mathbf{f}(\mathbf{p}_{s,1}),
\mathbf{f}(\mathbf{p}_{s,2}),
\ldots,
\mathbf{f}(\mathbf{p}_{s,M_s})
\right] \in \mathbb{C}^{M_{s,t} \times M_s}.
\end{equation}
Similarly, the FRV of the $n$-th reflecting element of the RIS at position
 $\mathbf{r}_{n} = [x_n, y_n]^T$
is given by $\mathbf{f}(\mathbf{r}_{s,n}) =
\left[
e^{j \frac{2\pi}{\lambda} \chi(\mathbf{r}_{n}, \phi_e^{1}, \phi_a^{1})},
\ldots,
e^{j \frac{2\pi}{\lambda} \chi(\mathbf{r}_{n}, \phi_e^{M_{s,r}}, \phi_a^{M_{s,r}})}
\right]^T \in \mathbb{C}^{M_{s,r} \times 1}$,
where $\frac{2\pi}{\lambda}\chi(\mathbf{r}_{n}, \phi_e^{j}, \phi_a^{j})$, for $1 \leq j \leq M_r$, denotes the phase difference of the $j$-th receive path between the position $\mathbf{r}_{n}$ and the reference origin of the RIS. Accordingly, the FRM at the RIS partition $p$ is given by
\begin{equation}
\mathbf{F}_\text{RIS}(\mathbf{r}_s) =
\left[
\mathbf{f}(\mathbf{r}_{s,1}),
\mathbf{f}(\mathbf{r}_{s,2}),
\ldots,
\mathbf{f}(\mathbf{r}_{s,N_p})
\right] \in \mathbb{C}^{M_{s,r} \times N_p}.
\end{equation}

Let $\boldsymbol{\Lambda}_s \in \mathbb{C}^{M_{s,r} \times M_{s,t}}$ denote the path response matrix for the link between the $s$-th sub-array and RIS partition $p$. The channel from the $s$-th sub-array to the RIS partition $p$ can then be written as
\begin{equation}
\mathbf{H}_{s,p}
= \mathbf{F}_\text{RIS}(\mathbf{r}_s)^H \boldsymbol{\Lambda}_s \mathbf{F}_\text{BS}(\mathbf p_s)
\in \mathbb{C}^{N_p \times M_s}.
\end{equation}

Next, let the channel from the RIS partition $p$ to the $d$-th user be denoted by $\mathbf{h}_{d} \in \mathbb{C}^{N_p \times 1}$. The resulting end-to-end equivalent channel from the $s$-th BS sub-array to the $d$-th user via RIS partition $p$ can be expressed as
\begin{equation}
\mathbf{g}_{s,p,d}^H(\mathbf{p}, \boldsymbol{\rho})
= \mathbf{h}_{d}^H \boldsymbol{\Phi}_p \mathbf{H}_{s,p}
\in \mathbb{C}^{1 \times M_s},
\end{equation}
where $\boldsymbol{\Phi}_p = \mathrm{diag}\left( e^{j\vartheta_1}, e^{j\vartheta_2}, \ldots, e^{j\vartheta_{N_p}} \right) \in \mathbb{C}^{N_p \times N_p}$ denotes the reflection matrix of the RIS partition $p$ and $\vartheta_n$ is the phase shift of  element $n$.

\subsection{Signal Model}
Suppose that $y_1 \sim \mathcal N_{\mathbb{C}}(0,1)$ is the signal transmitted from sub-array $1$ and $y_2 \sim \mathcal N_{\mathbb{C}}(0,1)$ is the signal transmitted from sub-array $2$. The  signal received at users $1$ and $2$ in  RIS partitions $1$ and $2$ can be expressed, respectively, as
\begin{align}\label{r_original1} \nonumber 
    r_1=&\big(\mathbf{h}_{1}^H \boldsymbol{\Phi}_1 \mathbf{H}_{1,1}+\mathbf{h}_{1}^H \boldsymbol{\Phi}_2 \mathbf{H}_{1,2}\big)\mathbf{w}_{1}y_1\\&+{\big(\mathbf{h}_{1}^H \boldsymbol{\Phi}_1 \mathbf{H}_{2,1}+\mathbf{h}_{1}^H \boldsymbol{\Phi}_2 \mathbf{H}_{2,2}\big)}\mathbf{w}_{2}y_2+n_1,
\end{align}
\begin{align}\label{r_original2} \nonumber 
r_2=&\big(\mathbf{h}_{2}^H \boldsymbol{\Phi}_1 \mathbf{H}_{2,1}+\mathbf{h}_{2}^H \boldsymbol{\Phi}_2 \mathbf{H}_{2,2}\big)\mathbf{w}_{2}y_2\\&+{\big(\mathbf{h}_{2}^H \boldsymbol{\Phi}_1 \mathbf{H}_{1,1}+\mathbf{h}_{2}^H \boldsymbol{\Phi}_2 \mathbf{H}_{1,2}\big)}\mathbf{w}_{1}y_1+n_2,
\end{align}
where $\mathbf{w}_{s} \in \mathbb{C}^{M_s \times 1}$, $s=\{1,2\}$ are the beamforming vectors of the sub-arrays and $n_d \backsim \mathcal N_{\mathbb C}(0, \sigma^2_d)$ is the receiver noise, where $\sigma^2_d$ denotes the noise power and $d=\{1,2\}$. Thus, the signal-to-interference plus noise ratio (SINR) for the $s$-th  sub-array at the BS to the $d$-th user in the presence of interference from the  sub-array $s^\prime$ is given by  
\begin{equation}\label{SINR_Exact}
    \begin{aligned}
        \gamma_{s,d}=  & \frac{\overbrace{\left| \mathbf{g}_{s,1,d}^H(\mathbf{p}_s, \boldsymbol{\rho}) \mathbf{w}_{s} \right|^2}^\textbf{aligned signal from partition 1}+\overbrace{\left| \mathbf{g}_{s,2,d}^H(\mathbf{p}_s, \boldsymbol{\rho}) \mathbf{w}_{s} \right|^2}^\textbf{non-aligned signal from partition 2}}{\underbrace{\sum_{p=1}^2 \left| \mathbf{g}_{s^\prime,p,d}^H(\mathbf{p}_{s^\prime}, \boldsymbol{\rho}) \mathbf{w}_{s'} \right|^2}_\textbf{interfering signals from the different sub-array over p=1, 2}+\sigma_d^2} \\ \nonumber &
        = \frac{\sum_{p=1}^2\left| \mathbf{g}_{s,p,d}^H(\mathbf{p}_s, \boldsymbol{\rho}) \mathbf{w}_{s} \right|^2}{\sum_{p=1}^2 \left| \mathbf{g}_{s^\prime,p,d}^H(\mathbf{p}_{s^\prime}, \boldsymbol{\rho}) \mathbf{w}_{s'} \right|^2+\sigma_d^2}.
    \end{aligned}
\end{equation}
Consequently, the achievable rate is  given by $R_{s,d}=\text{log}_2(1+\gamma_{s,d})$.

\section{MA-RIS Partitioning Design}
In this section, we design the framework of MA sub-array positions, their beamforming, and RIS partitioning.  
\subsection{Problem Formulation}
 
We aim to maximize the sum-rate of the signals transmitted from the aligned sub-arrays $1$ and $2$ to their corresponding users $1$ and $2$, respectively, while mitigating the interference caused by non-aligned sub-arrays. Thus,  the sum-rate of the considered system is $R=R_{1,1}+R_{2,2}$. An important outcome of this design is the reduction of inter-user interference alongside the enhancement of the desired signal power. To this end, each RIS partition is configured to form a directional beam toward its intended user. Accordingly, the sum-rate can be maximized by jointly optimizing the sub-array positions $\mathbf{P} = [\mathbf{p}_s, \mathbf{p}_{s'}]$, the beamforming matrix $\mathbf{W} = [\mathbf{w}_s, \mathbf{w}_{s'}]$, and the RIS partitioning vector $\boldsymbol{\rho}$, such that
\begin{subequations}\nonumber
\label{main_opt_program}
\begin{align}
&\mathcal P: ~\max_{\mathbf P, \mathbf{W},0 \preceq \boldsymbol{\rho} \preceq 1 } ~R
\label{main_opt_program_obj}\\
 & {\rm s.t.\ } ~\text{C$_{1}$:} ~\|\mathbf{p}_{s,m} - \mathbf{p}_{s,m'}\|_2 \geq \mathtt D, ~~~~~\forall m,m' \in \mathcal M_s, \label{main_opt_program_C1}\\
 &~~~~~~\text{C$_{2}$:} ~ \mathbf{p}_{s} \in \mathcal{C}_s, ~~~~~~~~~~s=\{1,2\}, \label{main_opt_program_C2}\\
& ~~~~~~\text{C$_3$:}~   \|\mathbf{w}_{s}\|^2+\|\mathbf{w}_{s'}\|^2 \leq P,\label{main_opt_program_C3}\\
&  ~~~~~~\text{C$_4$:}~ \sum_{p=1}^{2}\rho_p \leq 1,\label{main_opt_program_C4} 
\end{align}
\end{subequations} 
where $P$ and $\mathtt D$ are the available power budget and  the
minimum separation distance between the MAs of the same sub-array, respectively. Constraint \text{C$_1$} ensures sufficient separation among MAs of the BS sub-arrays to prevent mutual coupling, while constraint \text{C$_2$} restricts the MA positions to lie within the region $\mathcal C_s$. Constraint \text{C$_3$} represents the transmit power limit in the BS. Constraint C$_4$ ensures that the total allocated RIS portion does not exceed unity, implying that the number of assigned RIS elements cannot exceed the total number of available RIS elements. 

\subsection{Proposed Solution}
The strong coupling among the optimization variables  $\mathbf P, \mathbf{W},\boldsymbol{\rho}$ renders
problem $\mathcal P$ highly non-convex and intractable. To tackle this non-convexity, we propose an alternating optimization strategy that decomposes $\mathcal P$ into three subproblems, which are solved in an iterative manner as summarized in Algorithm 1.

1) \textit{ZF-based Beamforming Optimization:}
Under fixed values of $\mathbf P$ and $\boldsymbol{\rho}$, the optimization problem
associated with the design of $\mathbf{W}$ is written as follows 
\begin{subequations}\nonumber
\label{main_opt_program}
\begin{align}
&\mathcal P_1: ~\max_{\mathbf{W}} ~R
\label{main_opt_program_obj}\\
 & {\rm s.t.\ } ~\text{C$_3$:}~   \|\mathbf{w}_{s}\|^2+\|\mathbf{w}_{s^\prime}\|^2 \leq P.\label{main_opt_program_C3}
\end{align}
\end{subequations} 
Let $\mathbf{V} \in \mathbb{C}^{M \times 2}$ denote the communication channel from the BS to the users $d, d^\prime$ and defined as $\mathbf{V} = [\mathbf{g}_s(\mathbf p_s, \boldsymbol{\rho}),  \mathbf{g}_{s^\prime}(\mathbf p_{s^\prime}, \boldsymbol{\rho})]$, where $\mathbf{g}_s(\mathbf p_s, \boldsymbol{\rho})=
\begin{bmatrix}
\mathbf{g}_{s,p,d}(\mathbf p_s, \boldsymbol{\rho})\\
\mathbf{g}_{s^\prime, p^\prime, d}(\mathbf p_s, \boldsymbol{\rho})
\end{bmatrix}$ and $\mathbf{g}_{s'}(\mathbf p_{s^\prime}, \boldsymbol{\rho})=
\begin{bmatrix}
\mathbf{g}_{s,p,d^\prime}(\mathbf p_{s^\prime}, \boldsymbol{\rho})\\
\mathbf{g}_{s^\prime, p^\prime, d^\prime}(\mathbf p_{s^\prime}, \boldsymbol{\rho})
\end{bmatrix}$\footnote{We do not explicitly optimize the RIS phase shifts, as the optimal phase configuration can be obtained in closed form by aligning each RIS element phase $\vartheta_n$ with the phases of the corresponding BS-RIS and RIS-user channels under perfect channel state information (CSI)   \cite{8811733}. Consequently, the RIS suppresses the reflections associated with non-aligned links. Therefore, the contributions of the non-aligned beamformed signals, namely $\mathbf{g}_{s,p^\prime,d}(\mathbf p_s,\boldsymbol{\rho})\mathbf{w}_s$ and $\mathbf{g}_{s^\prime,p,d^\prime}(\mathbf p_{s^\prime},\boldsymbol{\rho})\mathbf{w}_{s^\prime}$, are assumed to be negligible and are omitted from the analysis.}. Let $\mathbf{V}^\dagger$ denote the Moore--Penrose pseudo-inverse of $\mathbf{V}$ such that
 $\mathbf{V}^\dagger \mathbf{V} = \mathbf{I}$, where $\mathbf{I}$ is the $2 \times 2$ identity matrix. Then, the following orthogonality condition holds:
\begin{equation}\label{ort}
\mathbf{v}_s^H \mathbf{g}_{s,p,d}(\mathbf p_s, \boldsymbol{\rho})) =
\begin{cases}
1, & s = p = d, \\
0, & \text{otherwise},
\end{cases}
\end{equation}
where $\mathbf{v}_s^H$ is the $s$-th row of $\mathbf{V}^\dagger$, for $s = 1,2$. Consequently, the beamforming direction for the $d$-th user is given by $\frac{\mathbf{v}_s}{\|\mathbf{v}_s\|_2}$, and the corresponding beamformer can be expressed as
\begin{equation}\label{beam}
\mathbf{w}_s = \sqrt{P_s}\, \frac{\mathbf{v}_s}{\|\mathbf{v}_s\|_2},
\end{equation}
where $P_s$ denotes the transmit power allocated to the $s$-th sub-array. Now, let   $\mathtt {P}=[P_{1}, P_{2}]$ be the power allocation decision for a
given $\mathbf{V}$; using \eref{ort} and \eref{beam}, the optimization problem
$\mathcal P_1$ can be re-written as
\begin{subequations}\nonumber
\label{main_opt_program}
\begin{align}
&\mathcal P_2: ~\min_{\mathtt{P}} ~-\sum_{s=1}^2\text{log}_2\big(1+\frac{P_s}{\|\mathbf{v}_s\|^2_2\sigma_s^2}\big)
\label{main_opt_program_obj}\\
 & {\rm s.t.\ }   ~\text{C$_1^2$:}~ \sum_{s=1}^2P_s \leq P.\label{main_opt_program_C3}
\end{align}
\end{subequations} 
The problem $\mathcal P_2$ can now be reformulated as an unconstrained program with the Karush–Kuhn–Tucker (KKT) conditions. The corresponding Lagrangian function can be expressed as
\begin{align} \label{16}
\mathcal L(\mathtt P, \Gamma)= \sum_{s=1}^2 - \text{log}_2(1+\frac{P_s}{\|\mathbf{v}_s\|^2_2\sigma_s^2})+\Gamma \big( \sum_{s=1}^2P_s-P\big),
\end{align}
where $\Gamma$ is the Lagrangian multiplier and the inequality in \text{C$_1^2$} is replaced by equality. By taking the partial derivative to \eref{16} and make it equal to zero, the optimum solution satisfies the following KKT conditions is given as
\begin{align} \label{pd}
P_s=
\left[
\frac{1}{\Gamma \ln 2}-
\|\mathbf{v}_s\|_2^2 \sigma_s^2
\right]^+,
\end{align}  
where  $[\xi]^+ \triangleq \max\{0, \xi\}$. The value of $\Gamma$ satisfies the   condition:
 $\sum_{s=1}^{2}
\left[
\frac{1}{\Gamma \ln(2)} - \|\mathbf{v}_s\|_2^2 \sigma_s^2
\right]^+ = P$.


2) \textit{RIS Partitioning Optimization:} Under fixed values of $\mathbf P$ and $\mathbf W$, the optimization problem
associated with the design of RIS partitioning
is formulated as follows 
\begin{subequations} \nonumber
\label{main_opt_program}
\begin{align}
\mathcal{P}_3: \quad 
&\max_{\rho_1,\rho_2} \quad R \\
\text{s.t.} \quad 
&\text{C}_4: \ \rho_1 + \rho_2 \leq 1, \\
&\text{C}_5: \ 0 \leq \rho_d \leq 1, \quad d=1,2.
\end{align}
\end{subequations}
$\mathcal{P}_3$ can be reduced to a one-dimensional optimization problem by substituting $\rho_2 = 1 - \rho_1$, resulting in
\begin{subequations}\nonumber
\label{reduced_problem}
\begin{align}
\mathcal{P}_4: \quad 
&\max_{\rho_1} \quad R \\
\text{s.t.} \quad 
& 0 \leq \rho_1 \leq 1.
\end{align}
\end{subequations}
The reduced problem $\mathcal{P}_4$ can be efficiently solved using a one-dimensional search. Specifically, the interval $[0,1]$ is discretized with a step size $\Delta$, and the achievable rate is evaluated for each candidate value of $\rho_1$. The optimal allocation is obtained as
\begin{equation}\label{rho}
\rho_1^\star = \arg\max_{0 \leq \rho_1 \leq 1} R,
\end{equation}
and $\rho_2^\star = 1 - \rho_1^\star$.

3) \textit{MA Position Optimization:} For given $\mathbf W$ and $\boldsymbol{\rho}$,
the optimization problem to determine the MA positions $\mathbf P$ can be formulated as follows
\begin{subequations}\nonumber
\label{main_opt_program}
\begin{align}
&\mathcal P_5: ~\max_{\mathbf P} ~R
\label{main_opt_program_obj}\\
 & {\rm s.t.\ } ~\text{C$_{1}$:} ~ \|\mathbf{p}_{s,m} - \mathbf{p}_{s,m'}\|_2 \geq \mathtt D, \forall m,m' \in \mathcal M_s, \label{main_opt_program_C1}\\
 &~~~~~~\text{C$_{2}$:} ~ \mathbf{p}_s \in \mathcal{C}_s, s=\{1,2 \}. \label{main_opt_program_C2}
\end{align}
\end{subequations} 
It is worth highlighting that $\mathcal P_5$ is computationally challenging since $\mathbf H_{s,p}$ is a nonlinear function of $\mathbf p_s$ involving
complex exponential functions. Additionally, the MA position variables $\{\mathbf p_{s,m} \}_{m=1, s=1,2}^{M_s}$ are jointly coupled. Hence, we solve $\mathcal P_5$ by leveraging the block coordinate descent (BCD) method. Specifically, in each BCD iteration, we optimize the position of the $m$-th transmit MA while keeping the positions of the other MAs $\{\mathbf p_{s,m'} \}_{m'=1, m' \neq m, s=1,2}^{M_s}$ fixed. This procedure is repeated until a convergence criterion is satisfied. 

Constraint C$_1$ in $\mathcal{P}_5$ is non-convex because it imposes a lower bound on the convex function $\|\mathbf{p}_{s,m}-\mathbf{p}_{s,m'}\|_2$. To address this challenge, we employ its first-order lower-bound approximation, given by
\begin{align}\label{conv}
  \frac{1}{\|\mathbf{p}^{(r)}_{s,m} - \mathbf{p}_{s,m'}\|_2} \big( \mathbf{p}^{(r)}_{s,m} - \mathbf{p}_{s,m'} \big)^T (\mathbf{p}_{s,m} - \mathbf{p}_{s,m'})\geq \mathtt D,  
\end{align}
where $\mathbf{p}^{(r)}_{s,m}$ denotes the value of $\mathbf{p}_{s,m}$ at the $r$-th iteration. It is easy to derive \eref{conv} as follows. First, the first-order lower bound of the convex function 
$\|\mathbf{p}_{s,m} - \mathbf{p}_{s,m'}\|_2$ at $\mathbf{p}_{s,m}^{(r)}$ is given by
\begin{align}
\|\mathbf{p}_{s,m} - \mathbf{p}_{s,m'}\|_2 
\geq & 
\|\mathbf{p}_{s,m}^{(r)} - \mathbf{p}_{s,m'}\|_2 
\\& \nonumber + \nabla \big(\|\mathbf{p}_{s,m}^{(r)} - \mathbf{p}_{s,m'}\|_2\big)^T
(\mathbf{p}_{s,m} - \mathbf{p}_{s,m^\prime}^{(r)}),
\end{align}
where the gradient term is
\begin{equation}
\nabla \big(\|\mathbf{p}_{s,m}^{(r)} - \mathbf{p}_{s,m'}\|_2\big)
= \frac{\mathbf{p}_{s,m}^{(r)} - \mathbf{p}_{s,m'}}{\|\mathbf{p}_{s,m}^{(r)} - \mathbf{p}_{m'}\|_2}
\end{equation}
Thus, we obtain
\begin{equation}\label{rhs}
\begin{aligned}
\|\mathbf{p}_{s,m} - & \mathbf{p}_{s,m'}\|_2 
\geq 
\|\mathbf{p}_{s,m}^{(r)} - \mathbf{p}_{s,m'}\|_2 \\
&\quad + \frac{1}{\|\mathbf{p}_{s,m}^{(r)} - \mathbf{p}_{s,m'}\|_2}
(\mathbf{p}_{s,m}^{(r)} - \mathbf{p}_{m'})^T(\mathbf{p}_{s,m} - \mathbf{p}_{s,m}^{(r)}).
\end{aligned}
\end{equation}
By simplifying the right-hand side of \eref{rhs}, we can easily get  \eref{conv}. Thus, the optimization problem for updating the MAs positions is as follows
\begin{subequations}\nonumber
\label{main_opt_program}
\begin{align}
&\mathcal P_6: ~\max_{\mathbf P} ~R
\label{main_opt_program_obj}\\
 & {\rm s.t.\ } ~\text{C$_{1}^6$:} ~ \frac{\big( \mathbf{p}^{(r)}_{s,m} - \mathbf{p}_{s,m'} \big)^T}{\|\mathbf{p}^{(r)}_{s,m} - \mathbf{p}_{s,m'}\|_2}  (\mathbf{p}_m - \mathbf{p}_{m'})\geq \mathtt D  , \forall m,m' \in \mathcal M_s,\label{main_opt_program_C1}\\
 &~~~~~~\text{C$_{2}$:} ~ \mathbf{p}_s \in \mathcal{C}_s, s=\{1,2\} \label{main_opt_program_C2}
\end{align}
\end{subequations}
which results in a convex optimization problem that is efficiently solvable using CVX with the MOSEK solver.

\begin{algorithm}[t!]
	\caption{Joint MA-RIS Partitioning Optimization}
	\label{alg_all}
	\begin{algorithmic}[1]
		\State   \textbf{Input:} user/RIS/BS locations; sub-array 2D site space $\mathcal{C}_s$; minimum inter-sub-array distance $\mathtt D$; step size $\Delta$; maximum number of iterations $T$  
      \While{the algorithm has not converged and the iteration
index $\leq T$}
       
       \State Update the beamforming  $\mathbf{W}$ using \eref{beam} and \eref{pd};
       \State Optimize $\boldsymbol{\rho}$  using $\mathcal P_4$ and \eref{rho};
     
        \State Update MA positions $\mathbf{P}$ by solving
$\mathcal P_6$ using BCD technique; 
    \EndWhile
	\end{algorithmic}
\end{algorithm} 
\subsection{Complexity Analysis}
The computational complexity of Algorithm~\ref{alg_all} is dominated by the iterative solutions of the underlying optimization subproblems $\mathcal{P}_4$ and $\mathcal{P}_6$. Specifically, $\mathcal{P}_4$ is solved via a one-dimensional search with complexity $\mathcal{O}(\epsilon_1)$, where $\epsilon_1$ denotes the number of discretization points. Meanwhile, $\mathcal{P}_6$ is addressed using BCD approach involving $M$ updates. Consequently, the complexity of solving $\mathcal{P}_6$ is $\mathcal{O}(\epsilon_2 M)$, where $\epsilon_2$ represents the number of BCD iterations required for convergence. Therefore, the overall computational complexity of Algorithm~\ref{alg_all} can be expressed as $\mathcal{O}\!\left(T\left(\epsilon_1+\epsilon_2 M\right)\right)$,  where $T$ denotes the number of outer iterations required for the alternating optimization algorithm to converge.

\section{Simulation Results}
In this section, the performance of the considered system
is evaluated through numerical simulations. The BS and RIS are located at coordinates $(0 ~\text{m},0 ~\text{m})$
and $(30~\text{m},~0 \text{m})$, respectively. Moreover, the users are randomly distributed within a circular region of radius $10$ m,
centered at $(30 ~\text{m}, -10~ \text{m})$. In addition, the planar far-field geometric channel model is adopted with $M_t=M_r=30$, where the elevation and azimuth angles of all propagation paths,  $\varphi_e^{i}, \varphi_a^{i}, \phi_e^{j}, \phi_a^{j}$, are uniformly distributed over $[0, \pi]$ \cite{asif2026channeluncertaintyawarerobustbeamforming}. LoS links are assumed to exist for both the BS–RIS and the RIS-user channels. Consequently, the diagonal elements of the BS–RIS path response matrix $\boldsymbol{\Lambda}_s$ are modeled as $\boldsymbol{\Lambda}_s[1,1] \sim \mathcal {CN}\bigg(0, \frac{\epsilon}{\epsilon+1} \mathcal P_0(\frac{d}{d_0})^{-\nu_1}\bigg)$ and $\boldsymbol{\Lambda}_s[i,i] \sim \mathcal {CN}\bigg(0, \frac{1}{\epsilon+1} \mathcal P_0(\frac{d}{d_0})^{-\nu_1}/(M_s-1)\bigg)$ for $i=2, \ldots, M_s$, where
$\epsilon$ and $\nu$ denote the Rician factor and the path-loss exponent, respectively, and $\mathcal P_0$ represents the average gain in channel power at a reference distance $d_0=1$ m. Moreover, RIS-user channels $\mathbf{h}_{d} \in \mathbb{C}^{N_p \times 1}$ are independent of MA 
positions and are modeled using Rician fading to account for
 LoS and non-LoS (NLoS) components, where
the path-loss exponent for the RIS-user link is denoted as $\nu_2$. Unless otherwise stated, the simulation parameters are set as follows: $\nu_1=2, \nu_2=2.5, \epsilon=3, \mathcal P_0= -30~ \text{dB}, \Delta=0.01, \mathcal C_1=[-\frac{A}{2}, \frac{A}{2}] \times [-\frac{A}{2}, \frac{A}{2}], \mathcal C_2=[-\frac{A}{2}+d_s, \frac{A}{2}+d_s] \times [-\frac{A}{2}+d_s, \frac{A}{2}+d_s],  A=3\lambda$ with $\lambda=0.1$ m, $d_s> A$, $\sigma_d=-80$ dBm, $\mathtt D=\lambda/2, P= 15$ Watts \cite{asif2026channeluncertaintyawarerobustbeamforming}.

We compare the performance of our proposed scheme of the joint optimization of MA positioning, beamforming, and RIS element selection, denoted as \textbf{Joint MA-RIS}, with the following schemes:
\begin{itemize}
    \item \textbf{Fixed Position Antenna:} In this scheme, we randomly select the beamforming and RIS elements.  
    \item \textbf{Random Position Antenna:} In this scheme, we randomly select the MA positions  and use the proposed scheme to optimize RIS beamforming and RIS elements.
    \item \textbf{Optimized MA locations with random beamforming and RIS element allocation (MA-Optimized):} In this scheme, we optimize  the MA positions  and randomly select RIS beamforming and RIS elements.
    
\end{itemize}

 \begin{figure}[t!]
    \centering  
    \includegraphics[width=0.94\linewidth]{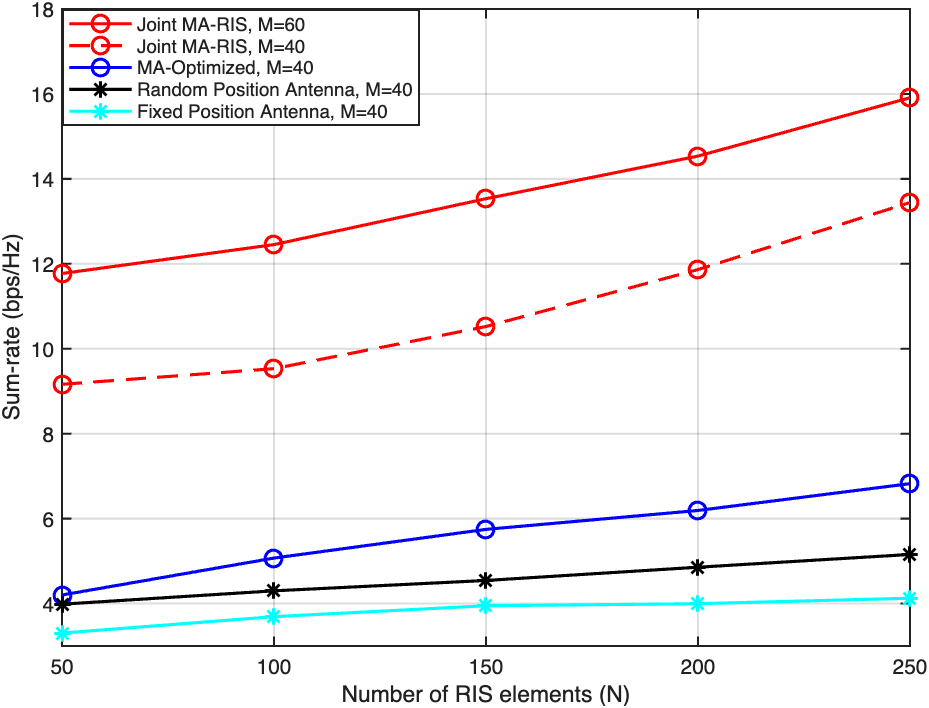}
    \caption{Sum-rate versus number of RIS elements $N$ for $2$ BS sub-arrays, $M=40, 60$, and $P=15$ Watt.}
    \label{fig5}
\end{figure}

Fig.~\ref{fig5} depicts the sum-rate versus the number of RIS elements $N$ for $D_{\mathrm{RIS}}=2$ and two RIS sub-arrays. As expected, the sum-rate increases with $N$ since a larger RIS provides higher passive beamforming gains and improves the quality of the RIS-assisted sub-array links. The advantage of the proposed joint optimization framework is evident across all values of $N$. In particular, the benefit of MA positioning optimization is validated by the superior performance of the MA-Optimized scheme compared to the fixed and random position antenna benchmarks. When beamforming and RIS element allocation are performed randomly, a considerable portion of the transmitted energy is directed toward unfavorable directions, resulting in a reduced sum-rate. In contrast, the proposed joint MA-RIS scheme jointly optimizes the MA positions, beamforming vectors, and RIS element allocation, thereby achieving the highest sum-rate among all considered schemes. The performance gains stem from the additional degrees of freedom offered by movable antennas, which enable the sub-arrays to be positioned at more favorable locations, enhancing the BS-RIS channels and improving beamforming efficiency. Although the MA-Optimized scheme optimizes antenna positions, its performance remains limited by the random beamforming and RIS allocation strategies. Overall, the results demonstrate that increasing the RIS size improves the performance of all schemes, while the proposed joint MA-RIS framework consistently provides the largest sum-rate gains.

\begin{figure}[!t]
    \centering  
    \includegraphics[width=0.95\linewidth]{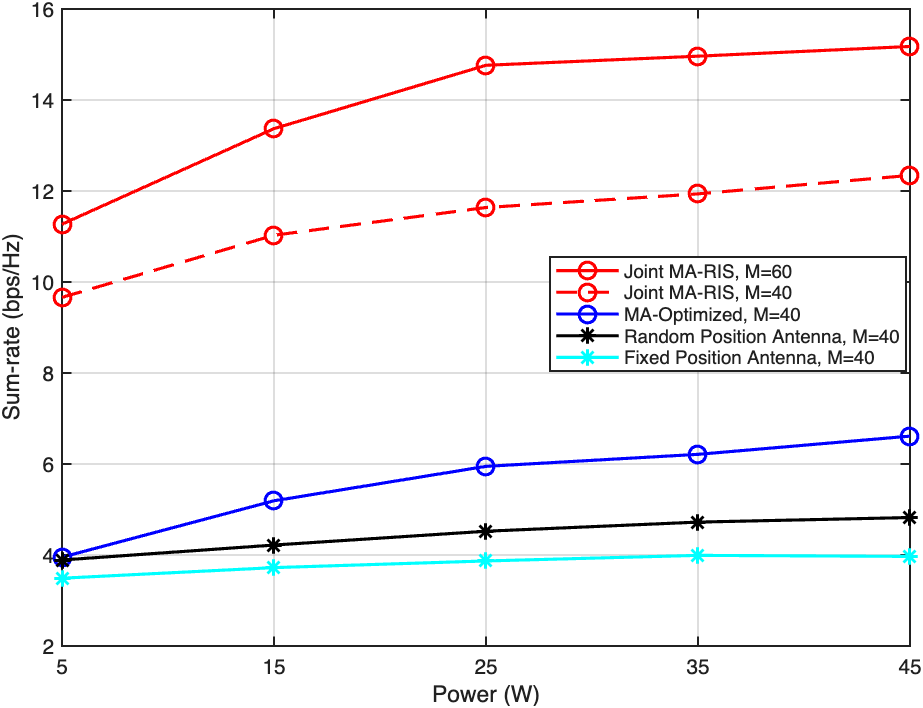}
    \caption{Sum-rate versus power-budget $P$ for $2$ BS sub-arrays, $N=100$, and $M=40, 60$.}
    \label{fig7}
\end{figure}

Fig.~\ref{fig7} illustrates the sum-rate of the system
versus the maximum available transmit power at the
BS. It is clear that the sum-rate increases with
the transmit power and the developed joint MA-RIS
framework provides remarkable improvement. Notably, as $M$ increases ($M=40, 60$), the sum-rate of the joint MA-RIS improves due to the additional spatial degrees of freedom available for beamforming and MA position optimization. Specifically, a larger number of MAs provides greater flexibility in adjusting the locations of the sub-array elements, enabling the BS to better adapt to the propagation environment and establish stronger BS-RIS channels.

\begin{figure}[!t]
    \centering  
    \includegraphics[width=0.95\linewidth]{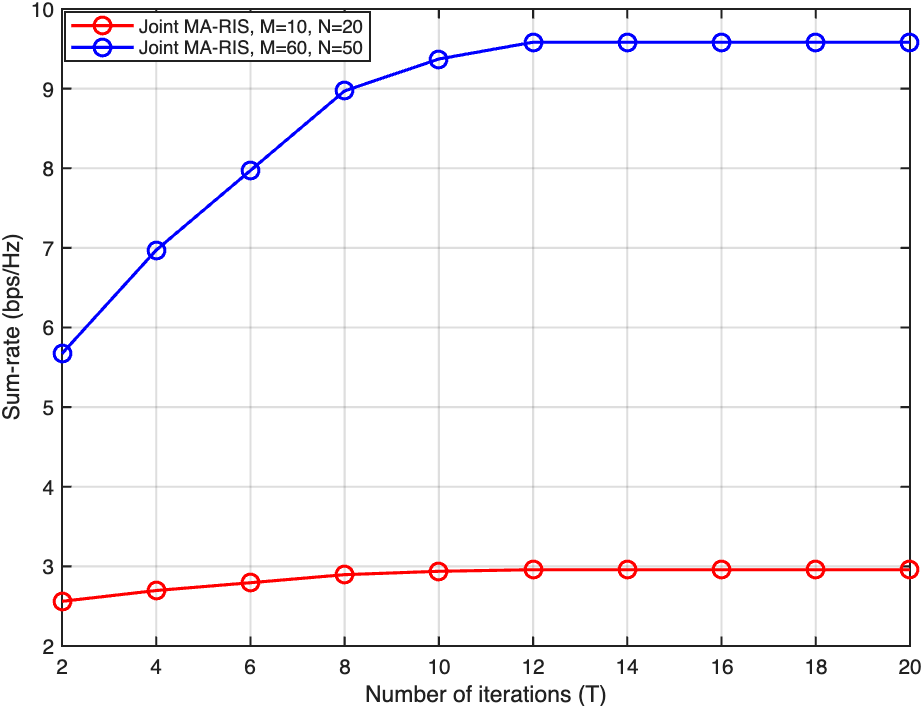}
    \caption{Sum-rate versus the number of iterations $T$ for $P=15$ W, $M=20, N=100$.}
    \label{fig8}
\end{figure}
 
Fig.~\ref{fig8} shows the convergence performance of the proposed optimization
framework with different system parameters: $M=10, N=20$ and $M=60, N=50$. The figure shows a stable convergence within a practical number of iterations. 

\section{Conclusion}
In this paper, we investigate a downlink RIS-assisted  system and proposed an optimization algorithm for MA positioning, sub-array beamforming, and RIS element selection. The proposed Joint MA-RIS scheme maximizes the sum-rate of the intended users while effectively mitigating interference by creating spatial notches in the directions of interfering signals and forming strong, directional beams toward the prospective users. Numerical results demonstrate that the proposed MA-RIS design outperforms baseline schemes such as MA-random, MA-optimized, and stationary   configuration. Concerning future work, we plan to extend the proposed framework to scenarios involving multiple RISs with joint phase-shift optimization. We also intend to investigate more practical channel models with imperfect CSI.  

\section*{Acknowledgment}
This research was supported by the Natural Sciences and Engineering Research Council of Canada (NSERC) through the Discovery Grants Program under Grant No. RGPIN-2026-06014.

\bibliographystyle{IEEEtran}
\bibliography{main-3}

\end{document}